\documentstyle[psfig,referee]{l-mn-mpa}

\newif\ifAMStwofonts
\AMStwofontstrue

\newcommand{\mv}{\mbox{$M_{V}$}}
\newcommand{\mi}{\mbox{$M_{I}$}}
\newcommand{\micl}{\mbox{$M_{I}^{\rm cl}$}}
\newcommand{\mbol}{\mbox{$M_{\rm bol}$}}

\newcommand{\ub}{\mbox{$U-B$}}
\newcommand{\bv}{\mbox{$B-V$}}
\newcommand{\vi}{\mbox{$V-I$}}
\newcommand{\dmo}{\mbox{$(m-M)_{0}$}}

\newcommand{\feh}{\mbox{[Fe/H]}}
\newcommand{\logt}{\mbox{$\log(t/{\rm yr})$}}
\newcommand{\Msun}{\mbox{$M_{\odot}$}}

\newcommand{\sub}[1]{\mbox{$_{\rm #1}$}}

\newcommand{\Mhef}{\mbox{$M\sub{Hef}$}}
\newcommand{\Mhe}{\mbox{$M\sub{cl}$}}
\newcommand{\Lhe}{\mbox{$L\sub{cl}$}}

\newcommand{\Mcore}{\mbox{$M\sub{c}$}}
\newcommand{\Menv}{\mbox{$M\sub{env}$}}
\newcommand{\Teff}{\mbox{$T\sub{eff}$}}
\newcommand{\logTe}{\mbox{$\log T\sub{eff}$}}
\newcommand{\logL}{\mbox{$\log(L/L_{\odot})$}}
\newcommand{\diff}{\mbox{d}}
\newcommand{\beq}{\begin{equation}}
\newcommand{\eeq}{\end{equation}}
\newcommand{\beqa}{\begin{eqnarray}}
\newcommand{\eeqa}{\end{eqnarray}}
\newcommand{\benu}{\begin{enumerate}}
\newcommand{\eenu}{\end{enumerate}}
\newcommand{\bite}{\begin{itemize}}
\newcommand{\eite}{\end{itemize}}
\newcommand{\bdes}{\begin{description}}
\newcommand{\edes}{\end{description}}
\newcommand{\refeq}[1]{equation (\protect\ref{#1})}
\newcommand{\reffig}[1]{Fig.\ \protect\ref{#1}}
\newcommand{\reftab}[1]{Table \protect\ref{#1}}
\newcommand{\refsec}[1]{Section \protect\ref{#1}}
\begin{document}

\title{A secondary clump of red giant stars: why and where}
\author{L\'eo Girardi}
\institute{Max-Planck-Institut f\"ur Astrophysik, 
	Karl-Schwarzschild-Str.\ 1, D-85740 Garching bei M\"unchen,
	Germany \\
	E-mail: leo@mpa-garching.mpg.de} 
\date{Accepted 19?? ???.
      Received 1998 ???;
      in original form 1998 ???}

\maketitle
\markboth{L.~Girardi: A secondary clump of red giants}
{L.~Girardi: A secondary clump of red giants}

\label{firstpage}
\thispagestyle{empty}
\mbox{\ }\clearpage\setcounter{page}{1}

\thispagestyle{empty}
\parbox[b]{15cm}{\mbox{~~~~} \\[4.0cm] 
\begin{abstract}
Based on the results of detailed population synthesis models, Girardi
et al.\ (1998) recently claimed that the clump of red giants in the
colour--magnitude diagram (CMD) of composite stellar populations
should present an extension to lower luminosities, which goes down to
about 0.4~mag below the main clump. This feature is made of stars just
massive enough for having ignited helium in non-degenerate conditions,
and therefore corresponds to a limited interval of stellar masses and
ages. In the present models, which include moderate convective
overshooting, it corresponds to $\sim1$~Gyr old populations.

In this paper, we go into more details about the origin and properties
of this feature. We first compare the clump theoretical models with
data for clusters of different ages and metallicities, basically
confirming the predicted behaviours.  We then refine the previous
models in order to show that: (i) The faint extension is expected to
be clearly separated from the main clump in the CMD of metal-rich
populations, defining a `secondary clump' by itself. (ii) It should be
present in all galactic fields containing $\sim1$~Gyr old stars and
with mean metallicities higher than about $Z=0.004$. (iii) It should
be particularly strong, if compared to the main red clump, in galaxies
which have increased their star formation rate in the last Gyr or so
of their evolution. In fact, secondary clumps similar to the model
predictions are observed in the CMD of nearby stars from {\em
Hipparcos} data, and in those of some Large Magellanic Cloud fields
observed to date.  There are also several reasons why this secondary
clump may be missing or hidden in other observed CMDs of galaxy
fields. For instance, it becomes undistinguishable from the main clump
if the photometric errors or differential absorption are larger than
about 0.2~mag.  Nonetheless, this structure may provide important
constraints to the star formation history of Local Group galaxies. We
comment also on the intrinsic luminosity variation and dispersion of
clump stars, which may limit their use as either absolute or relative
distance indicators, respectively.
\end{abstract}

}

\thispagestyle{empty}
\mbox{\ }\clearpage\setcounter{page}{1}

\section{Introduction}
\label{sec_intro}

Since the works by Cannon (1970) and Faulkner \& Cannon (1973), the
clump of red giants in the colour--magnitude diagram (CMD) of
intermediate-age and old open clusters is recognised as being formed
by stars in the stage of central helium burning (CHeB). The near
constancy of the clump absolute magnitude in these clusters was
correctly interpreted as the result of He-ignition in a
electron-degenerate core. Under these conditions, He-burning can not
start until the stellar core mass attains a critical value of about
0.45~\Msun. It then follows that all low-mass stars (i.e.\ those which
develop a degenerate He-core after H-exhaustion) have similar core
masses at the beginning of He-burning, and hence similar luminosities.

The red giant clump is also a remarkable feature in the CMD of
composite stellar populations, like the fields of nearby galaxies.
Since the clumps of all stellar populations older than about 1~Gyr
fall in the same region of the CMD, they define a feature usually as
striking to the eye as the main sequence. Recently, the {\em
Hipparcos} data allowed to clearly see the clump in the CMD of nearby
stars (Perryman et al.\ 1997).

Due to its origin and the nearly constancy in luminosity, the red
clump has been usually considered to contain few information about its
parent population. Its use as a distance indicator has been proposed
by Cannon (1970), and since then applied by several authors, like
e.g.\ Seidel, Da Costa \& Demarque (1987), Hatzidimitriou \& Hawkins
(1989) and others. It has been recently re-proposed by Paczy\'nski \&
Stanek (1998), following the release of the {\em Hipparcos}
catalogue. Indeed, the mean absolute magnitude of the clump defined by
{\em Hipparcos} can be measured with a precision of hundredths of
magnitude, and used as a reference for \dmo\ determinations. The
apparent constancy of the $I_0$ magnitude with the \vi\ colour inside
the clump, in different stellar systems, was considered as evidence
that the mean clump absolute magnitude \micl\ does not depend on
metallicity (Paczy\'nski \& Stanek 1998; Stanek \& Garnavich 1998;
Udalski et al.\ 1998; Stanek, Zaritsky \& Harris 1998).  This
conclusion, however, has been questioned by Girardi et al.\ (1998;
hereafter Paper~I). They made use of a large set of evolutionary
tracks to show that more massive clump stars are systematically bluer
than the less massive ones, so that the colour range spanned by the
{\em Hipparcos} clump is largely caused by the dispersion of masses
(and hence ages) of local stars, and not simply by their metallicity.
In this case, the constancy of the mean clump $I$-band magnitude with
colour can be simply reproduced by models, and understood as the
result of age segregation inside the clump. \micl, instead, might be
systematically lower (i.e.\ brighter) for lower metallicities. The
model predictions then can be used to provide a first-order correction
to distance determinations derived from clump stars (see also Cole
1998).  For the LMC, the corrections are such that
$\dmo=18.28\pm0.14$~mag as estimated in Paper~I, or
$\dmo=18.36\pm0.17$~mag cf.\ Cole (1998).

Another prediction of Paper~I is that the stars just massive enough
for igniting He under non-degenerate conditions, should define a {\em
secondary clumpy feature}, located about 0.3~mag below the clump of
the lower-mass stars, and at its blue extremity. A feature similar to
the predicted one was found in the CMD from {\em Hipparcos}.

In this paper, we go into some important details of the theoretical
predictions raised in Paper~I. First, we briefly recall the underlying
stellar evolution theory (\refsec{sec_theory}). Then we compare the
theoretical behaviour of clump stars with data from stellar clusters
of different ages and metallicities (\refsec{sec_clusters}).  This
comparison gives support to the behaviour predicted by models. We
present simple models for the clump structure in composite stellar
populations (\refsec{sec_galaxies}), which predict the appearance of
the secondary clump. The conditions for its presence in CMDs of
different galaxy fields are examined. We suggest the association of
this theoretically predicted feature with some observational ones, and
comment on the impact of these results in the interpretation of data
for the Magellanic Clouds and other Local Group
galaxies. \refsec{sec_conclu} summarises the main conclusions.

\section{Brief review of the theory}
\label{sec_theory}

In stars of mass higher than a given value,
$M\ge\Mhef\sim2-2.5$~\Msun\ (see e.g.\ Chiosi, Bertelli \& Bressan
1992), helium ignition takes place under non-degenerate conditions,
when the stellar centre attains a critical value of 
temperature and density. These are the intermediate- and high mass
stars.  In this case the core mass at helium ignition is a simple
monotonically increasing function of stellar mass. The same behaviour
holds for the luminosity of the stars during the subsequent core
helium burning (CHeB) phase.

In low-mass stars, i.e.\ those with $M<\Mhef$, an electron-degenerate
core forms after the central hydrogen exhaustion. Electron conduction
reduces the temperature gradient in the core, and the efficient
neutrino cooling at the centre makes the temperature maximum to shift
to an off-centre region. In this case, the critical temperatures and
densities for helium ignition are attained only later in the
evolution, when the core mass has grown to a value of about
$\Mcore\simeq0.45$~\Msun. This value is nearly constant for the stars
with mass lower than about $\Mhef-0.3$~\Msun, depending only little on
chemical composition and stellar mass. As a result of their similar
core masses \Mcore, the CHeB phase occurs at similar luminosities for
all low-mass stars.  However, the helium burning stars of higher
masses are slightly hotter than those of lower masses, due to their
different envelope masses \Menv. Higher effective temperatures, and
slightly lower luminosities, are also reached at the limit of very low
masses, for which $\Menv\rightarrow0$.

	\begin{figure}
	\psfig{file=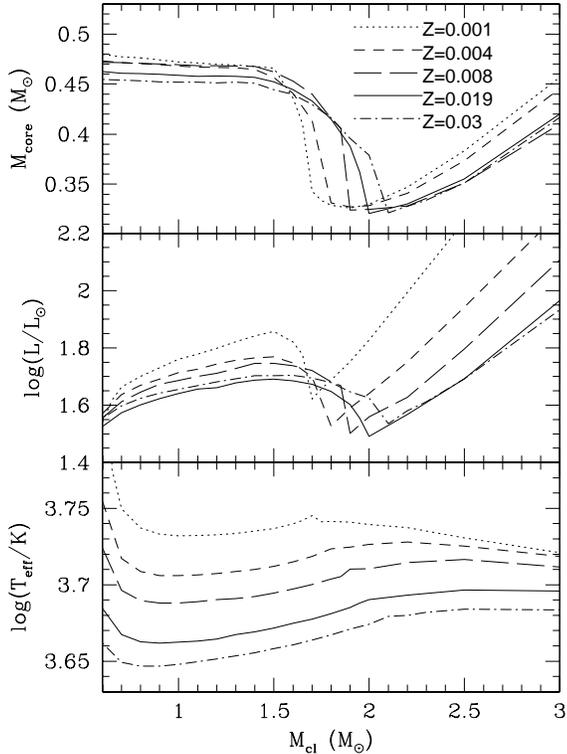,width=8.3cm}
        \caption{Stellar parameters at the first stage of quiescent
helium burning (i.e.\ the ZAHB for low-mass stars, or the CHeB stage
of lowest luminosity for intermediate-mass ones), as a function of
stellar mass and for 5 values of initial metallicity, as derived from
Girardi et al.\ (1999) models. They are: the core mass $\Mcore(M)$ in
the upper panel, $\log L(M)$ in the middle, and $\log\Teff(M)$ in the
lower one. }
	\label{fig_mcore}
	\end{figure} 

These behaviours are illustrated in \reffig{fig_mcore}. We plot the
values of \Mcore, the luminosity $L$ and effective temperature \Teff\
at the initial stage of quiescent CHeB, from the Girardi et al.\
(1999) stellar models.  This point of the stellar tracks coincides
with the `zero-age horizontal branch' (ZAHB) usually defined for
low-mass stars, and with the one of minimum luminosity attained
during central He-burning for intermediate-mass stars.  Some basic
characteristics of these stellar tracks have been already described in
Girardi \& Bertelli (1998) and in Paper~I. 
Suffice it to recall here that they include a
moderate amount of convective overshooting from stellar cores.

\begin{table}
\caption{The transition masses \Mhef\ and their corresponding ages
$t(\Mhef)$.}
\label{tab_mhef}
\begin{tabular}{lllll}
\noalign{\smallskip}\hline\noalign{\smallskip}
$Z$ & $Y$ & overshoot & $\Mhef/\Msun$ & $t(\Mhef)$/Gyr \\
\noalign{\smallskip}\hline\noalign{\smallskip}
0.001 & 0.230 & moderate & 1.7 & 1.22 \\
0.004 & 0.240 & moderate & 1.8 & 1.13 \\
0.008 & 0.250 & moderate & 1.9 & 1.09 \\
0.019 & 0.273 & moderate & 2.0 & 1.13 \\
0.030 & 0.300 & moderate & 2.1 & 1.27 \\
0.019 & 0.273 & no       & 2.4 & 0.54 \\
\noalign{\smallskip}\hline\noalign{\smallskip}
\end{tabular}
\end{table}

To be noticed in \reffig{fig_mcore} is that both $\Mcore(M)$ and
$L(M)$ assume a minimum value for $M\sim2$~\Msun, of
$\Mcore\sim0.33$~\Msun\ and $\logL\simeq1.5-1.6$. The luminosity
minimum is used to define the transition masses \Mhef\ referred to in
the present work.  Their values are indicated in Table~\ref{tab_mhef}
for 5 different values of metallicity. The main-sequence lifetime
$t(\Mhef)$ of the corresponding star is also presented. One can notice
that whereas \Mhef\ decreases systematically with the metallicity $Z$,
the corresponding transition age $t(\Mhef)$ is nearly constant (see
also Sweigart, Greggio \& Renzini 1990 for a detailed discussion of
this point), and close to 1.1~Gyr. The \Mhef\ and $t(\Mhef)$ values we
find are typical of models computed with moderate convective
overshoot.  Additionally, the table presents the same quantities as
derived from a set of solar metallicity tracks in which the classical
Schwarzschild criterion for convection (i.e.\ no overshooting) is
adopted. In this case, \Mhef\ is about 20 percent larger, whereas
$t(\Mhef)$ about 50 percent lower, than in the case with moderate
overshooting.

It is worth recalling that the behaviours presented in
\reffig{fig_mcore} are not exclusive to the Girardi et al.\ (1999) set
of evolutionary tracks. Instead, they are common to models computed
either with or without convective overshooting, and regardless of the
scheme adopted to deal with the `breathing pulses of convection'
during the late CHeB evolution. The reader is referred to the works by
Becker \& Iben (1980), Maeder \& Meynet (1989), Sweigart et al.\
(1990), Castellani, Chieffi \& Straniero (1992), for similar
descriptions of the function $\Mcore(M)$, which are however based on
very different sets of evolutionary tracks. Castellani et al.\ (1992)
also present predictions for $L(M)$ which are similar to the ones
described by the present models. Only the stellar models adopting the
approximation that \Mcore\ is constant for all low-mass ZAHB stars
(e.g.\ Seidel, Demarque \& Weinberg 1987) probably do not follow all
the trends shown in \reffig{fig_mcore}.

	\begin{figure}
	\psfig{file=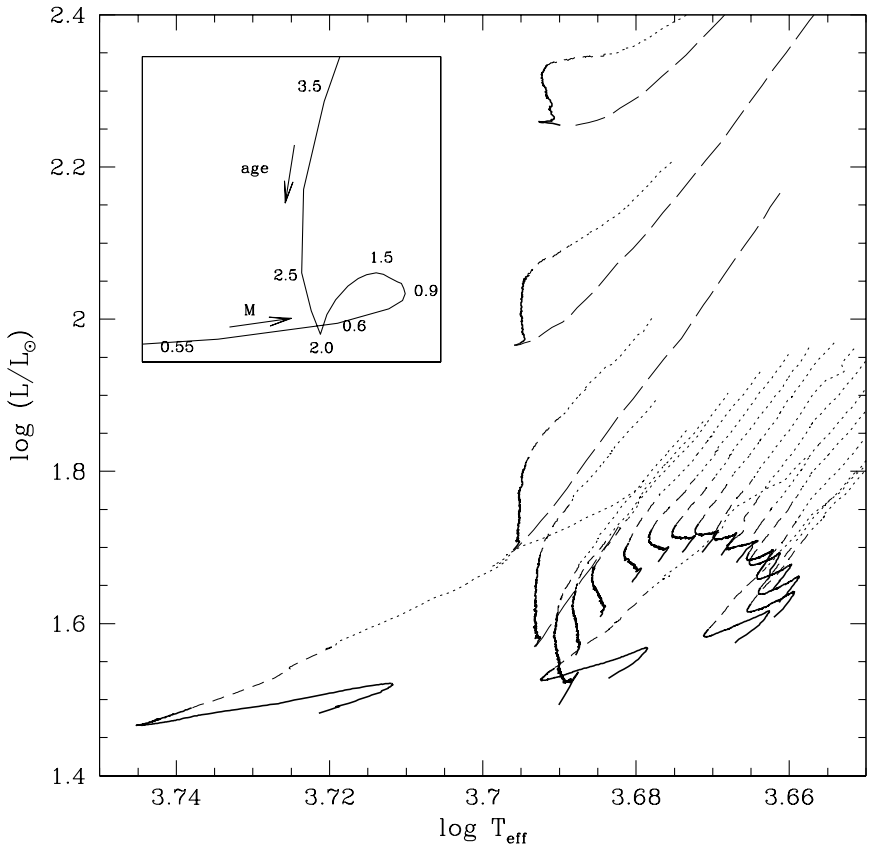,width=12cm}
        \caption{Evolution of $Z=0.019$ tracks on the HR diagram
during the stage of quiescent CHeB, for several mass values.  The
solid lines denote the evolution during most of the CHeB lifetime: for
low-mass stars, they start when the gravitational contraction provides
less than 1 percent of the stellar luminosity (it corresponds to the
stage of ZAHB), whereas for intermediate-mass stars they begin at the
CHeB stage of lowest luminosity.  In both cases, the lines end when 70
percent of the CHeB lifetime $t\sub{He}$ is reached. Thus, these solid
lines correspond to most of the clump red giants (or to the horizontal
branch in the limit of lowest masses).  Dashed and dotted lines
illustrate phases of faster evolution during the CHeB. The
short-dashed lines denote the evolution from 70 percent up to 85
percent of $t\sub{He}$, whereas the dotted ones go up to 99 percent of
$t\sub{He}$. Moreover, the long-dashed lines present the initial stage
of quiescent CHeB of intermediate-mass stars; in the mass range here
plotted ($M\le3.5$~\Msun), this phase never exceeds 12 percent of
$t\sub{He}$. The insert shows schematically how the stellar masses and
ages vary along the sequence of CHeB models. Some representative mass
values (in \Msun) are indicated along this line. }
	\label{fig_cheb}
	\end{figure} 

Another important aspect is that the individual stellar tracks, in the
mass and metallicity ranges here considered, change little their
characteristics during most of the CHeB evolution. This is illustrated
in \reffig{fig_cheb}, for the case of solar-metallicity ($Z=0.019$)
CHeB models. In low-mass tracks of $M>0.8$~\Msun, 70 percent of the
CHeB lifetime, $t\sub{He}$, is spent within a box of $\Delta\log
L<0.12$~dex (or $\Delta\mbol<0.3$~mag) by $\Delta\logTe<0.006$~dex in
the HR diagram. Only at the late stages of CHeB evolution the tracks
deviate significantly from their ZAHB positions. A similar situation
occurs for intermediate-mass stars, for which an initial phase of fast
contraction is followed by a phase of much slower evolution in the HR
diagram. As a useful first approximation, we can then consider the
locus of these slow-evolution phases in the HR diagram as
representative of the behaviour of the complete CHeB tracks of
different masses.  Finally, we remark that the excursions
in temperature during the CHeB phase increase at lower metallicities
and masses, but become appreciably higher than the above-mentioned
limits only for the tracks with $Z<0.004$ or with $M<0.8$~\Msun.

From Figs.~\ref{fig_mcore} and \ref{fig_cheb} (see also figure~1 in
Paper~I), two clear sequences can be noticed:
	\bite
	\item
	For $M<\Mhef=2$~\Msun, and going from higher to lower masses,
clump stars have almost constant luminosities at decreasing
temperatures.  At $M\la0.9$~\Msun, this sequence bends towards much
hotter temperatures.  The less luminous stars in this mass range are
both the most massive (due to their lower core masses) and the less
massive ones.
	\item
	For $M>\Mhef$, and going from lower to higher masses, clump
stars have increasing luminosity and almost-constant temperatures. The
minimum luminosity, for $M=\Mhef$, is about 0.4~mag lower than that of
stars with slightly lower mass.
	\eite

\section{The red clump in clusters}
\label{sec_clusters}

The behaviour predicted by the models can be compared to that found in
star clusters of different ages and metallicities. The data for
clusters in the Magellanic Clouds are particularly useful because
their distances and reddening can be considered as constant in a first
approximation. Therefore, they allow to infer how the clump luminosity
changes as a function of the cluster parameters.  Galactic open
clusters instead generally suffer from a higher uncertainty in their
relative distance modulus.

\subsection{The red clump luminosity}
\label{sec_clump_brightness}

\begin{figure}
	\psfig{file=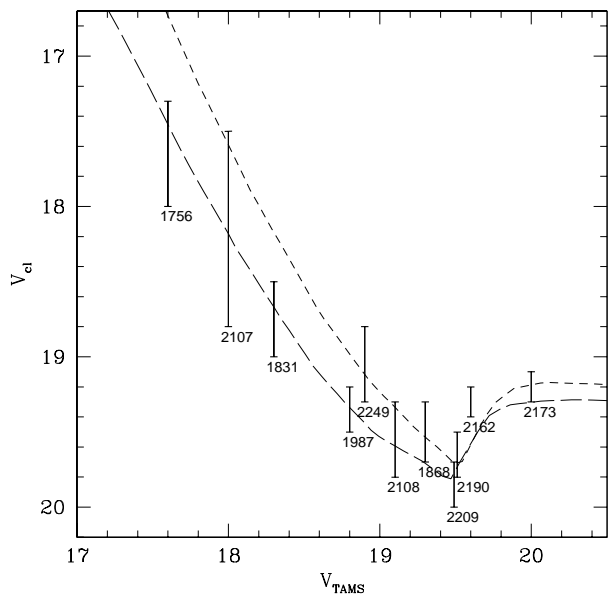,width=8.3cm}
        \caption{The behaviour of the clump magnitude $V_{\rm cl}$ as
a function of the main sequence termination magnitude $V_{\rm TAMS}$
in LMC clusters, as observed by Corsi et al.\ (1994), and as predicted
by the Girardi et al.\ (1999) models. The vertical lines link the the
mean clump position, with that of its lower extremity (see data in
table~17 of Corsi et al.\ 1994). The NGC numbers of each cluster are
indicated.  The ages range from about 0.1 to 2 Gyr, from left to
right. Sequences of CHeB models (at the stage of initial CHeB, or
ZAHB) for $Z=0.008$ and 0.004 are plotted adopting an apparent
distance modulus of 18.5~mag. These model lines have the same meaning
as indicated in \reffig{fig_mcore}.  }
	\label{fig_corsi}
	\end{figure}

Corsi et al.\ (1994) collected $BV$ data for a dozen LMC clusters,
with ages ranging from about 0.2 to 1.2 Gyr, aiming at locating the
precise age interval in which the RGB develops in stellar
populations. One of the best diagnostics of the development of the RGB
is in fact given by the behaviour of the clump luminosity: it
increases when the RGB appears and becomes almost constant
afterwards. Figure~\ref{fig_corsi} presents the data from Corsi et
al.\ (1994, their table~17) -- namely the clump magnitude $V_{\rm cl}$
against that of the termination of the main sequence, $V_{\rm TAMS}$,
for each cluster. The vertical lines actually give the possible range
of magnitude for the base of the clump, i.e.\ the one ranging between
the mean measured clump, and the faintest clump stars observed in each
cluster ($\left<V_{Cl}\right>$ and $V_{Cl,m}$, respectively, in Corsi
et al.\ 1994 tables). Superimposed, are the relations predicted by
Girardi et al.\ (1999) models for metallicities $Z=0.004$ and $0.008$,
at the stage of the lowest luminosity during the CHeB (cf.\ also
\reffig{fig_mcore}). The reader should keep in mind that, in the
models, the mean clump magnitude is typically located $\la0.15$~mag
above these lines.

The models describe well the behaviour suggested by the data,
especially the presence of a minimum in the $V_{\rm cl}(V_{\rm TAMS})$
relation, for $V_{\rm TAMS}\simeq19.5$, corresponding to the clusters
NGC~1868, NGC~2108, NGC~2190 and NGC~2209. Only clusters to the right
of this minimum (i.e.\ older and presumably containing stars with
electron-degenerate helium cores), like NGC~2162 and NGC~2173 are
found to undoubtedly contain RGB stars (see Corsi et al.\ 1994; and
the comments in Girardi \& Bertelli 1998). Of course, different
assumptions for the apparent distance modulus of the clusters, or
different ways of defining their main sequence termination magnitudes,
could change a little the presentation of the data and models in this
diagram.

Other recent and homogeneous compilation of clump magnitudes is
provided by Udalski (1998b). They collected $VI$ photometry for
several intermediate-age and old Magellanic Cloud clusters (with ages
from about 1.5 to 12 Gyr), concluding that the clump magnitude
$I_0^{\rm cl}$ does not change appreciably in this age range. Together
with the results of Udalski (1998a), these data were used to argue in
favour of the mean $I$-band clump magnitude as a standard candle, and
of a low value of the LMC distance modulus -- namely
$18.18\pm0.06$~mag (see also Udalski et al.\ 1998; Stanek et al.\
1998).

	\begin{figure}
	\psfig{file=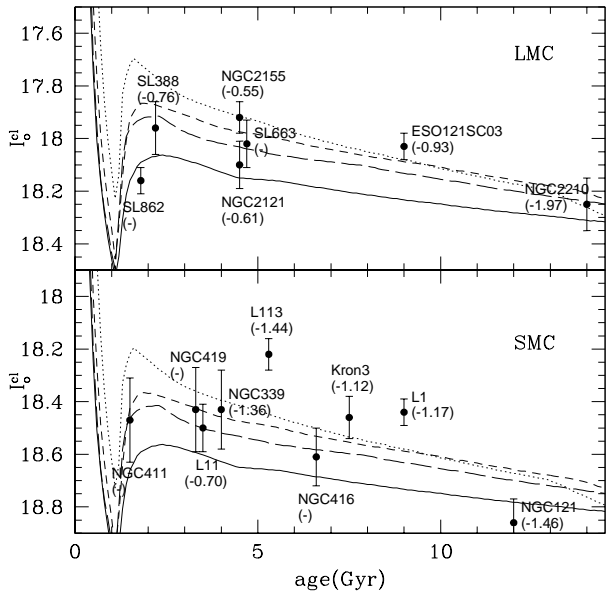,width=8.3cm}
        \caption{The behaviour of the clump magnitude as a function of
age, in the range from 2 to about 10 Gyr, in both the LMC (upper
panel) and the SMC (lower one). The data by Udalski (1998b) are
compared to Girardi et al.\ (1999) models for several metallicities.
For each cluster, we indicate the value of \feh\ as measured either by
Olszewski et al.\ (1991; for the LMC) or by Da Costa \& Hatzidimitriou
(1998; for the SMC).  The model lines have the same meaning as in
\reffig{fig_mcore}: from above to below we have models with
metallicities $Z=0.001$, 0.004, 0.008, and 0.019, corresponding to
$\feh=-1.3$, $-0.7$, $-0.4$, and 0, respectively.
}
	\label{fig_udalski}
	\end{figure} 

Figure~\ref{fig_udalski} shows Udalski's (1998b, their tables~2, 3 and
4) data for clusters in both Clouds, compared to the results of our
models with different metallicities.  In plotting the data, we have
not applied geometric corrections to the distance modulus of each
cluster, as adopted by Udalski. Also, we did not
attempt to correct the cluster ages by him presented, in order to put
them into a single and consistent age scale. We should keep in mind
that age errors as large as 1~Gyr may be present in the data.

The data points present a non-negligible scatter in $I_0^{\rm cl}$,
which can not be exclusively attributed to the observational
errors. The only clear trend in the data is that the two clusters
older than 10~Gyr present clumps fainter by $0.25-0.3$~mag with
respect to the younger ones.

In \reffig{fig_udalski} we plot also the model predictions for the
mean clump absolute magnitude \micl\ for several metallicities, as
determined by properly averaging the luminosity of CHeB stars along
the theoretical isochrones. Models were shifted
by a constant distance modulus of $\dmo=18.3$~mag for the LMC, and
18.7~mag for the SMC, in order to fit the appropriate range of the
data.

It is clear that in the age range of $\ga2$~Gyr, the models predict
that the $I$-band clump magnitude mildly decreases with age, and
mildly increases with the metallicity.  In order to better compare
data and models, we need to consider the age-metallicity relation
intrinsic to the cluster data. It is well known that different methods
may lead to discrepant metallicity estimates for the same clusters. In
the following, we make use only of the metallicity determinations by
Olszewski et al.\ (1991) for the LMC clusters, and Da Costa \&
Hatzidimitriou (1998) for the SMC ones, whose values are also
indicated in \reffig{fig_udalski}. These works made use of the same
spectroscopic method for deriving abundance values. Therefore the
metallicity scale should be rather homogeneous.  For the sake of this
homogeneity, we also use the SMC data as expressed in the Zinn \& West
(1984) scale (see figure~4 in Da Costa \& Hatzidimitriou 1998).

For the LMC, the behaviour of the clump magnitude with age indicated
by the data seems to be well described by the present models. The only
exception is the cluster ESO121\,SC03, which apparently presents a too
bright clump if compared to our lowest metallicity models. The
discrepancy is however not larger than 0.1~mag. Moreover, we should
keep in mind the unique characteristics of this cluster (see Mateo,
Hodge \& Schommer 1986): it is located very far from the LMC disk, and
is the only known cluster in the LMC with age comprised between 4~Gyr
and the classical old clusters with $>12$~Gyr (see Da Costa 1991;
Olszewski, Suntzeff \& Mateo 1996; Geisler et al.\ 1997). Therefore,
it is not clear whether it could really be considered as a member of
the LMC disk, and whether it is located at the same distance of the
bulk of LMC clusters.

In the SMC, the situation clearly changes. The dispersion of clump
magnitudes increases for the clusters in this galaxy, becoming larger
than that suggested by the models of different metallicities. What is
remarkable is the extremely bright clump of L~113 ($\sim0.2$~mag
brighter than the clusters of similar age), and the faint clump of the
cluster NGC~121. In the case of L~113, part of the observed difference
could be understood as the result of the apparently anomalous (and
low) metallicity of this cluster (see the careful discussion about the
SMC age--metallicity relation by Da Costa \& Hatzidimitriou 1998),
which according to the models would imply a brighter clump.  However,
the explanation in terms of metallicity differences only would require
a much larger dependence of the clump absolute magnitude on
metallicity than predicted by the present models. Moreover, a clump so
bright is not observed in the case of NGC~339, despite its similar
metallicity and age (notice however the large error bar in this latter
case). Maybe the key to understanding the clump magnitudes of L~113
and NGC~121, is that these objects are really located at different
distances along the line-of-sight. If this is the case for most of the
SMC clusters, the comparison with theoretical models under the
assumption of a single distance modulus would lose significance.

We conclude that {\em no firm conclusion about the behaviour of the
mean clump absolute magnitude with age and metallicity, for ages
larger than 2 Gyr, can be based on the data for so few clusters}. The
main problem with LMC data is essentially the lack of bona-fide
clusters at the LMC distance with ages between 4 and 12~Gyr. In the
case of the SMC, depth effects probably complicate very much the
analysis.

On the other hand, we do not find in these data the evidences that
model over-estimate the \micl\ dependence on either age or
metallicity, as suggested by Udalski (1998ab). In
\refsec{sec_distance} we further comment on this aspect. We recall
that Udalski's conclusions for the SMC clusters were largely based on
a data sample for which empirical corrections to the distances were
applied. These distance corrections are based on the observed
variation of the magnitude of RR Lyrae stars in the neighbouring
fields.  For clusters like L~1 and L~113, distance corrections of up
to $\pm0.2$~mag were derived by extrapolating the values derived for
other lines-of-sight.  We regard this procedure, when applied to
single clusters, as potential sources of error in the analysis.

\subsection{The red clump colour}

Another important observational relation is that given by
Hatzidimitriou (1991) between the cluster age and the $B\!-\!R$ colour
difference between the clump and the RGB, $d_{B\!-\!R}$.  The good
correlation between these quantities allowed her to suggest the use of
$d_{B\!-\!R}$ as an age indicator for intermediate-age and old star
clusters.

	\begin{figure}
	\psfig{file=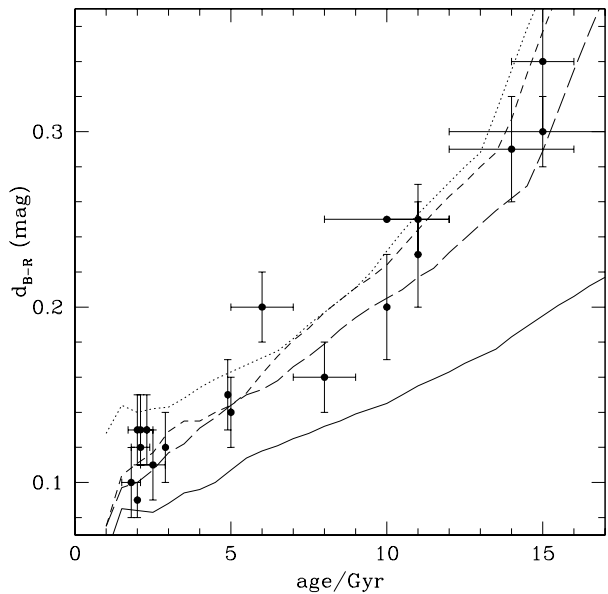,width=8.3cm}
        \caption{The behaviour of the colour difference between the
mean clump and the RGB at the same level in $R$, $d_{B\:-\:R}$, as a
function of the cluster age, according to the data collected by
Hatzidimitriou (1991, her table~1), and the Girardi et al.\ (1999)
models for several metallicities. The lines have the same meaning as
indicated in \reffig{fig_mcore}.
}
	\label{fig_hatzi}
	\end{figure} 

Hatzidimitriou's (1991) data sample is shown in
\reffig{fig_hatzi}. Notice that, in plotting these data, we have made
no effort in order to revise the ages of the individual clusters. In
fact, the oldest clusters in the sample should better be attributed
ages of $\sim12$~Gyr, instead of the 15~Gyr shown in the plot, as a
result of the recent changes in the absolute age scale of the old
globular clusters (see e.g.\ Gratton et al.\ 1997; Salaris \& Weiss
1997, 1998). In this contest, the age scale presented in
\reffig{fig_hatzi} should be considered as a relative one.

The present models for metallicities from $Z=0.001$ to $0.008$ seem to
follow a common $d_{B\!-\!R}$ vs.\ age relation, which reproduces well
the correlation found in the data. The same does not happen for the
$Z=0.019$ ones, which seem to predict too small $d_{B\!-\!R}$ values
at a given age. With respect to the data, we remark that only 3 of the
clusters plotted in \reffig{fig_hatzi} have metallicities comparable
to solar (i.e.\ with [Fe/H]$>-0.3$), the remaining ones being more
metal poor.

The reason for the $Z=0.019$ isochrones presenting so low values of
$d_{B\!-\!R}$, as compared with the more metal-poor ones, is that for
this value of metallicity the bump in the luminosity function along
the RGB happens to be located below the luminosity level of the clump.
Above this bump, the RGB evolutionary tracks shift to slightly higher
temperatures. Therefore, $d_{B\!-\!R}$ values get suddenly lower as we
go from the $Z=0.008$ to the $Z=0.019$ tracks. In the observed CMDs,
the reference values of the RGB colour at the level of the clump
probably do not take into account this subtle effect, since in these
cases the RGB ridge line is drawn by eye, connecting stars both below
and above the bump feature. A more detailed comparison with the data
would be worth in order to clarify if this effect may be responsible
for the apparent discrepancy in the $Z=0.019$ models. In the present
work, we limit ourselves to comment that {\em the observed trend of
$d_{B\!-\!R}$ with age is clearly present in the models}. If there is
any discrepancy with the observational data, this is in the sense that
our most metal-rich models {\em underestimate} this colour difference.

\section{The red clump in galaxies}
\label{sec_galaxies}

Once we have tested the model predictions about the clump position in
CMDs as a function of age and metallicity, by comparing them with
observations of star clusters, we can proceed our modeling of more
complex stellar populations.

\subsection{Basic theory}
\label{sec_mass_dist}

The red clump observed in the CMDs of galaxy fields may be described
by the sum of those deriving from single-burst stellar populations of
different ages and metallicities.  In the case of a composite
population of single metallicity, if we neglect mass-loss on the RGB,
the number $N(\Mhe)\diff\Mhe$ of clump stars with mass in the interval
$[\Mhe, \Mhe+\diff\Mhe]$, is expressed by
	\beq
N(\Mhe) \propto \phi_M(\Mhe)\, \psi[T-t(\Mhe)]\, t\sub{He}(\Mhe)
	\label{eq_clump}
	\eeq
(see Paper~I), where $\phi_M$ is the IMF, $\psi(T-t)$ is the SFR at
the epoch of stellar birth $T-t(\Mhe)$, and $t\sub{He}$ is the
lifetime of the CHeB phase.

	\begin{figure}
	\psfig{file=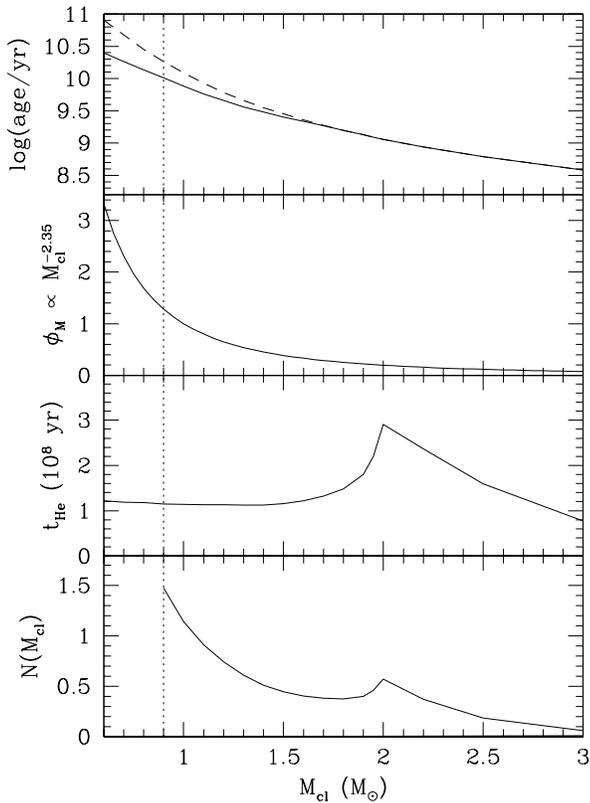,width=8.3cm}
        \caption{Quantities which enter into the determination of
the mass distribution of clump stars, as derived from $Z=0.008$
tracks.  In the panels from top to bottom, we have: (1) the age vs.\
mass relation, for models assuming mass-loss according to Reimers'
(1975) formula with efficiency $\eta=0.4$ (continuous line), and for
the alternative case of no mass-loss (dashed line); (2) the Salpeter
IMF, $\phi_M\propto M^{-2.35}$, in arbitrary units; (3) the lifetime
in the core He-burning phase; and finally (4) the resulting mass
distribution, $N(\Mhe)$, in arbitrary units, for the case of a
constant SFR during the complete age interval up to 10~Gyr. The
vertical dotted line represents the lower mass limit for clump stars
in a galaxy field of 10~Gyr and solar metallicity, in the case of
Reimers' mass-loss (see upper panel).}
	\label{fig_mass}
	\end{figure} 

Figure~\ref{fig_mass} shows the quantities which enter into this
equation (except for the SFR) as a function of \Mhe, and as derived
from our $Z=0.019$ models. To be noticed in the top panel is that the
age vs.\ mass relation for clump stars, $t(\Mhe)$, depends on the
amount of mass-loss taken place during the RGB phase. Hence, this
function is presented for both the cases (i) of no mass-loss, and (ii)
assuming a Reimers' (1975) mass-loss formula with efficiency
$\eta=0.4$ (see Renzini \& Fusi Pecci 1988). In the latter case, clump
stars of mass lower than 0.90~\Msun\ would not be present in stellar
populations younger than 10 Gyr. This gives an estimate for the lower
mass limit of clump stars of solar metallicity. This value would be
increased to 1.06~\Msun\ in the case of no mass-loss. Finally,
mass-loss does not affect significantly the stars with
$\Mhe\ga1.5$~\Msun.

The second panel from top to bottom presents the Salpeter IMF, with
its well-known steep increase for lower-mass stars. We notice that
also this function should be slightly corrected in order to consider
mass-loss. In fact, what should be used in the IMF of \refeq{eq_clump}
is the initial mass $M$ of the progenitor of a clump star with mass
\Mhe. $M-\Mhe$ is however always less than 0.2~\Msun\ for the case of
Reimers' mass-loss with $\eta=0.4$\footnote{It can be noticed from the
upper panel of \reffig{fig_mass}: for a given age, the mass difference
between the two drawn lines (dashed minus continuous) gives a good
estimate of the mass lost by each star in the RGB, $M-\Mhe$.}.  Thus,
for the illustrative purposes we are interested in this moment, we
prefer to ignore this effect and simply adopt the Salpeter IMF in
\reffig{eq_clump}.

The third panel from top to bottom shows the CHeB lifetime as a
function of \Mhe. As remarked in Girardi \& Bertelli (1998) and
Paper~I, this lifetime increases for stars with lower core masses
\Mcore\ at the moment of He-ignition, i.e.\ those in the vicinity of
\Mhef. This is so because the lower the initial core mass in the CHeB
phase (see upper panel of \reffig{fig_mcore}), (i) the more nuclear
fuel becomes available for the H-burning shell, and (ii) the lower is
the luminosity at which CHeB takes place (middle panel of
\reffig{fig_mcore}).

Finally, the bottom panel of \reffig{fig_mass} presents the final mass
distribution $N(\Mhe)$ [from \refeq{eq_clump}] for a galaxy model in
which the SFR is assumed to be constant over the complete age interval
up to 10~Gyr. The maximum age of 10~Gyr implies the cut-off in this
distribution along the vertical line at $\Mhe=0.9$~\Msun. An important
aspect evidenced by this figure is that {\em the mass distribution of
clump stars has a second maximum at $M\simeq\Mhef=2$~\Msun}. Moreover,
{\em the CHeB stars with mass slightly higher than $\Mhef=2$~\Msun\
are not severely under-represented with respect to lower-mass
ones}. In the particular case here shown, the number ratio between
stars with $\Mhe\ge2$~\Msun\ and those with $2>(\Mhe/\Msun)>0.9$ is of
0.19.  Moreover, it can be noticed that clump stars with mass higher
than about $\Mhe\ga2.5$~\Msun\ become very few, due to the reduction
of both their lifetimes and representativeness in the IMF.

\subsection{The secondary clump in different galaxy models}
\label{sec_subclump}

The CHeB stars with mass close to \Mhef\ are not only relatively
frequent (bottom panel of \reffig{fig_mass}), but are also located
into a particular region of the HR diagram (see \reffig{fig_mcore}).
They are up to 0.4~mag fainter than the slightly less-massive and
older clump stars. Also, in metal-rich stellar populations they are
the bluest among clump stars. Due to these characteristics, they can
define a {\em secondary clump} in the CMD.

This feature was predicted in Paper~I, and identified with a small
group of stars in the {\em Hipparcos} CMD of nearby stars (see
Figs.~2, 3 and 4 therein). The reader is referred to that work for a
generic description of this point. Suffice it to recall here that this
feature corresponds to the approximate position of the $2$~\Msun\
track in \reffig{fig_cheb}. The main red clump would correspond to
CHeB stars in the mass range from about 1.9 to 0.9~\Msun.

	\begin{figure}
	\psfig{file=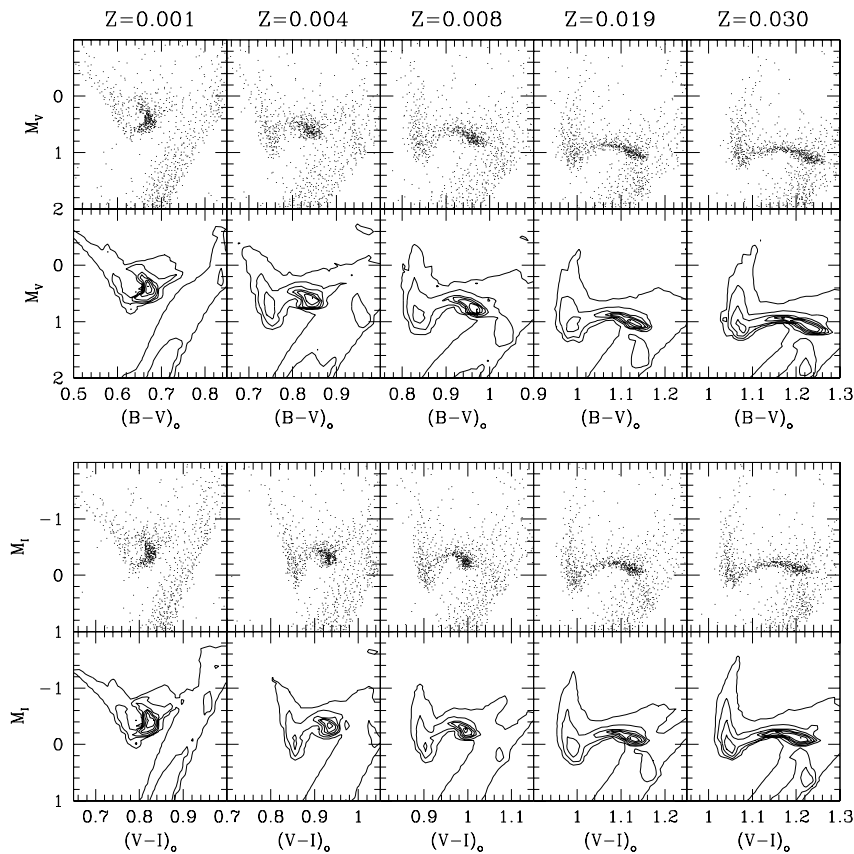,width=12cm}
        \caption{Synthetic CMDs showing the red clump region for
different metallicities, in the $BV$ (top panels) and $VI$ (bottom
ones) planes. Each simulated CMD is presented in two versions: in the
first one, 6 contour lines limit regions with the same stellar
density $N$, and are spaced at constant intervals of
$N^{1/2}$. Immediately above, we show the corresponding synthetic CMD
for a total of 1000 stars. Models assume constant SFR from 0.1 to 10
Gyr, constant mean metallicity, and a small metallicity dispersion
(see text).  Panels from left to right are for mean metallicities
$Z=0.001$, 0.004, 0.008, 0.019, and 0.03.  The most densely populated
region of the diagrams represent the main body of the red clump. To
its blue, we have the structure which contains both the faint
secondary clump and the plume of brighter clump stars. The RGB runs
slightly to the red of the main clump. It presents another
concentration of stars (more evident in the right-hand panels), which
corresponds to the RGB bump.}
	\label{fig_secclump}
	\end{figure} 

In \reffig{fig_secclump}, we make use of synthetic CMDs in order to
predict how the red clump looks at different metallicities. This in
the ideal cases in which the star formation has proceeded at a
constant rate and with a constant value of mean metallicity. The
models include also a small Gaussian dispersion of metallicities, of
$\sigma_Z/Z=0.1$, and the effect of mass loss along the RGB. The
method of synthesis here used is the same as in Paper~I, and takes
into account all stars in all evolutionary stages, with all ages and
metallicities, which can be contributing to each point of a given CMD.

The plots for the highest metallicities show red clumps covering a
considerable range in colour. This colour dispersion may be
interpreted as an age sequence, with younger (more massive) clump
stars having bluer colours. At the extreme left of this sequence, the
clump luminosity decreases by about 0.4~mag. This region of the CMD is
identified with the secondary clump we refer. It corresponds to a real
increase in the stellar density, as can be seen from the contour plots
in \reffig{fig_secclump}. Moreover, in \refsec{sec_massresol} below we
argue that this feature should become even denser (i.e.\ more clumpy)
if we further improve the mass resolution of our stellar models.

It is also clear from this plot that the red clump is expected to be
narrower in colour for lower metallicities. Moreover, in the diagrams
for $Z=0.001$ (which may well illustrate the typical case for
Pop.~II), the low-mass (and hence oldest) tail of the clump is already
bending to the blue, and then partially superimposes on the region
which would correspond to the secondary clump. As a consequence, the
latter becomes almost undistinguishable from the main clump. Both
could not be separated in real CMDs with some realistic distribution
of photometric errors. On the contrary, in stellar populations of
higher metallicity, the secondary clump clearly separates from the
main one.  We can conclude that {\em the secondary red clump is well
separated from the main one only at metallicities higher than about
$Z=0.004$.}  For lower metallicities, all clump stars with masses
higher than about 0.7~\Msun\ may be mixed into a single clump feature.

Of course, the case of constant SFR and metallicity is illustrative,
but not representative of most galaxy fields. In Paper~I we presented
a synthetic CMD of stars in the solar vicinity, in which an
age--metallicity relation was assumed, together with an age-dependent
factor to account for the fact that older stars are located at larger
heights above the galactic plane. Also in that case a secondary clump
was present in the models, and yet clearly separated from the main
clump.

	\begin{figure}
	\psfig{file=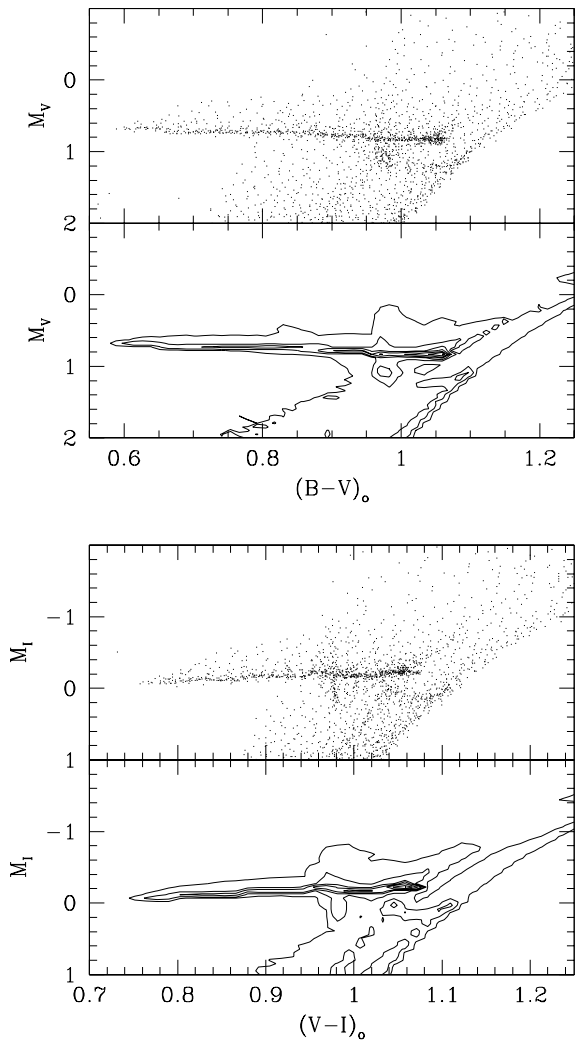,width=8.3cm}
        \caption{Synthetic CMDs showing the red clump for a galaxy
model which assumes a simple chemical evolution history, in the $BV$
and $VI$ planes. The different panels are as in
\reffig{fig_secclump}. The synthetic CMDs include a total number of
2000 stars. The model assume a exponentially decreasing SFR from 0.1
to 12 Gyr, and a metallicity linearly increasing with the galaxy age
(see the text for details).}
	\label{fig_secclump_amr}
	\end{figure} 

In \reffig{fig_secclump_amr} we show another simulation for a galaxy
following an age-metallicity relation. In this model, the SFR is
assumed to be exponentially decreasing with the galactic age, with an
e-folding time of 5~Gyr. Then we adopt the metallicity evolution
predicted by the simple closed-box model of chemical evolution (Searle
\& Sargent 1972): $Z$ increases linearly with time, from a minimum
value of $Z=0.001$ at 12~Gyr ago, to the present value of
$Z=0.019$. The result is that the main red clump becomes a feature
very extended in colour. The secondary red clump is however still
present, at a colour corresponding to that of the 1~Gyr old population
[ie.\ at $\mv\simeq1.1$, $(\bv)_0\simeq0.97$ in the top panels, and at
$\mi\simeq0.1$, $(\vi)_0\simeq0.98$ in the bottom panels], but
superimposed on the RGB of the more metal-poor populations. In an
observed CMD, such a feature could hardly be distinguished from the
background of RGB stars.  Moreover, we can also observe that other
faint concentrations of stars come out along the RGB and slightly
below the main clump; they correspond to the RGB bump of metal-rich
stellar populations, and should not be confused with the secondary
clump of CHeB stars. Finally, the reader is invited to compare these
synthetic CMDs with that of the M~31 field presented by Stanek \&
Garnavich (1998, their figures~2 and 4). Such a comparison evidences
the capability of the present models of reproducing the general
features of the M~31 clump, especially in regard to its colour range
and inclination in the \mi~vs.~\vi\ diagram.

The case of galaxy populations with large metallicity dispersions at a
given age are not presented here. The properties of the corresponding
clumps are however easy to derive from the above plots. Since the
secondary clump is in general less populated than the main one, a
large metallicity dispersion at 1~Gyr would make it to become a
feature of lower density, and therefore not easily distinguishable from
the background population of RGB stars present in most CMDs.

In the case of a constant SFR, the total number of low-mass clump
stars is basically determined by the maximum age $T$ of the stellar
populations we consider. For $T=10$~Gyr, we have a lower mass limit of
0.9~\Msun, which is increased to 1.3~\Msun\ if we adopt
$T=4$~Gyr. Thus, in the case of relatively young galaxies, the number
ratio between the stars with $\Mhe>\Mhef$ and those with $\Mhe<\Mhef$
increases (see e.g.\ the bottom panel of \reffig{fig_mass}). This
circumstance would make {\em the secondary clump more evident, when
compared to the main clump, in galaxies which have increased their SFR
in the last Gyr or so of their evolution.}

\subsection{On the mass range and resolution of the clump models}
\label{sec_massresol}

The secondary clump can also be interpreted as being the faint
extremity of a large vertical structure in the CMD, i.e.\ the sequence
of CHeB stars which ignited helium in non-degenerate conditions. The
brightest part of this sequence consists of a plume of stars departing
from the main red clump to higher luminosities. It has been noticed in
the CMD of several Local Group galaxies (e.g.\ Zaritsky \& Lin 1997;
Beaulieu \& Sacket 1998; Alves et al.\ 1998; Tolstoy et al.\ 1998),
and is often referred to as the `vertical red clump' or `blue loop
feature'. This feature is also evident in the simulations of
\reffig{fig_secclump}.

But how could the secondary clump have been missed in previous studies
of synthetic CMDs, contrarily to this less populated bright tail? A
crucial aspect here regards the mass resolution of the grids of
evolutionary tracks used to construct synthetic CMDs.  In Paper~I, we
made use of grids with a resolution of $\sim0.1$~\Msun\ in the
vicinity of \Mhef. These closely spaced tracks allowed the
construction of accurate isochrones even for this mass (and hence age)
interval in which the evolutionary features change so remarkably.

	\begin{figure}
	\psfig{file=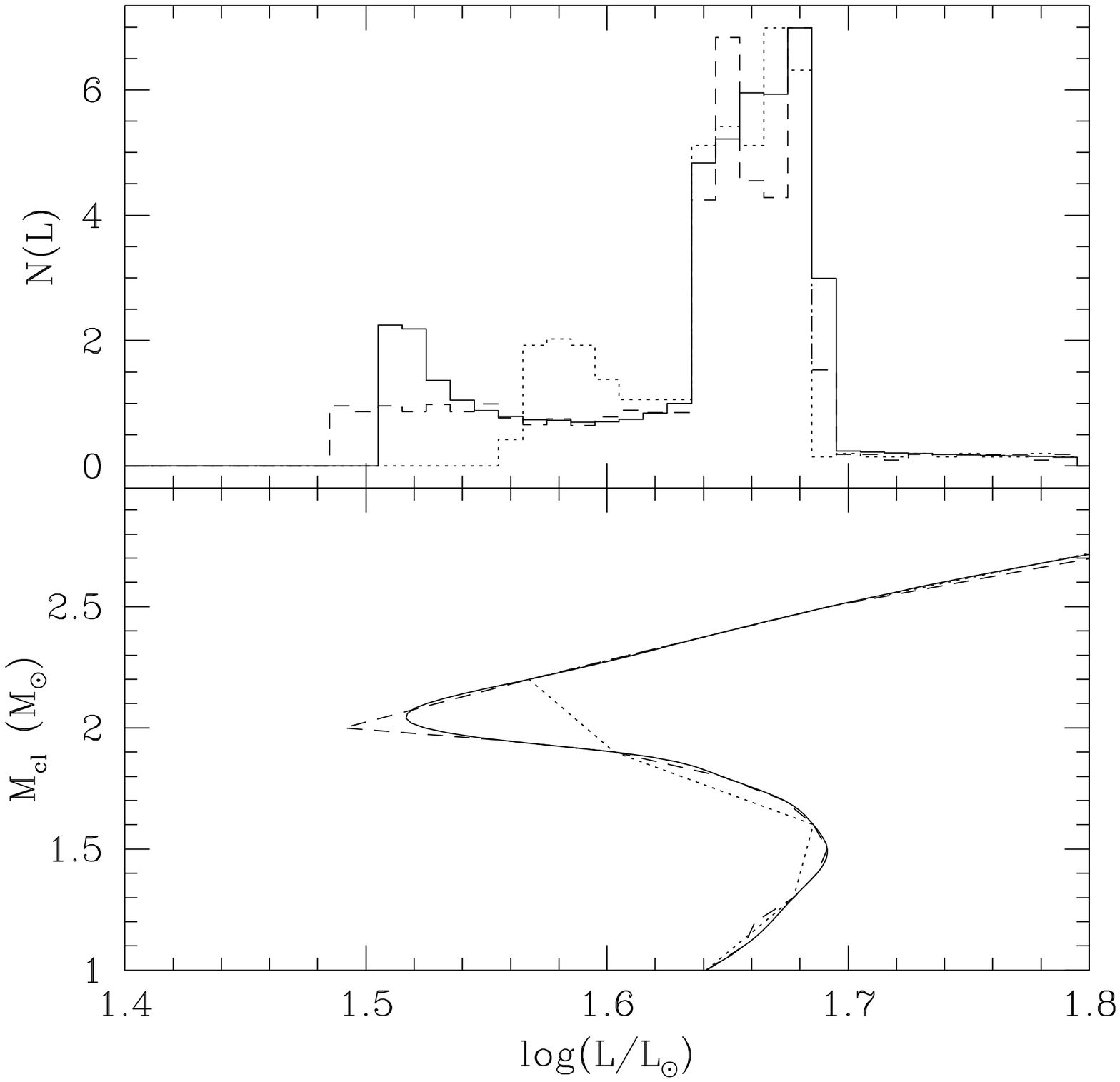,width=8.3cm}
        \caption{Lower panel: $\Lhe(\Mhe)$ relations as obtained from
grids of evolutionary tracks computed with different mass resolutions:
$\Delta \Mhe=0.3$~\Msun\ for the dotted line, $\Delta \Mhe=0.1$~\Msun\
(typically) for the dashed line, and the idealised case of
$\Delta\Mhe\rightarrow0$ for the continuous line. Upper panel: The
luminosity functions of clump stars as derived from these sequences,
in arbitrary units. Notice the different characteristics of the
fainter clump in the three cases.}
	\label{fig_lum_dist}
	\end{figure} 

To give a better idea of the importance of having a dense grid of
tracks in order to find the secondary clump, \reffig{fig_lum_dist}
shows simple simulations of the luminosity distribution obtained from
stellar models at the stage of lowest luminosity during the CHeB
(i.e.\ ignoring the remaining CHeB evolution), calculated with
different mass resolutions. In these simple models, \refeq{eq_clump}
gives the number of stars to be assigned to each mass bin $[\Mhe,
\Mhe+\Delta \Mhe]$, with a constant luminosity distribution over the
corresponding luminosity bin $[\Lhe, \Lhe+\Delta \Lhe]$ (see bottom
panel). The contributions from the stars in the complete mass interval
from 1 to 3~\Msun\ are summed in order to construct the luminosity
functions shown in the upper panel.

All the simulations in \reffig{fig_lum_dist} clearly show a main clump
at $1.64<\logL<1.70$ (upper panel), corresponding to the ZAHB stars
with $1.0<\Mhe/\Msun<1.8$ (see bottom panel), and a second feature
located at a somewhat fainter luminosity. In the case with the worst
mass resolution ($\Delta \Mhe=0.3$~\Msun; dotted line), the minimum of
luminosity for CHeB stars with $\Mhe=\Mhef$ is missed, and therefore
the resulting secondary clump is located only 0.06~dex in luminosity
($0.15$~mag) below the main one. If we then were to consider the
excursion of $\la0.3$~mag upwards in luminosity during the stellar
CHeB evolution in these models, the faint feature would be blurred
enough to be mixed with the main red clump.

However, as we improve the mass resolution of the models, the
secondary clump emerges at lower luminosities. The dashed line shows
the clump as derived from our CHeB sequences, for which $\Delta \Mhe$
is of typically 0.1~\Msun. In this case the faint extension of the red
clump becomes a more extended feature in the luminosity function,
defining a second maximum $\sim0.14$~dex ($0.35$~mag) below the main
one.  This maximum is however not very pronounced.  

The continuous line instead shows the ideal situation of a continuous
and smooth CHeB sequence (see bottom panel), in order to simulate the
$\Delta\Mhe\rightarrow0$ case. This line was obtained from a natural
cubic spline passing through most (but not all) of the computed model
points. In this case, the minimum of the $\Lhe(\Mhe)$ function defines
a caustic in the luminosity distribution, which causes the faint
secondary clump to become a quite striking feature, clearly separated
in luminosity from the main clump by a gap of 0.35~mag. In this case,
the excursions in luminosity during the CHeB evolution would not be
enough to blur this feature with the main red clump. 

Sweigart, Greggio \& Renzini (1989, 1990) computed evolutionary tracks
of hydrogen burning stars up to the He-ignition, with a very fine mass
resolution, of 0.05~\Msun\ close to \Mhef. These models are especially
suited to study the velocity at which the RGB appears in aging
stellar populations. The functions $\Mhe(M)$ they derive (Sweigart et
al.\ 1990) are indeed very smooth along the transition from low- to
intermediate-mass stars. Even if they do not compute the subsequent
He-burning evolution from these tracks, their $\Mhe(M)$ relations
clearly suggest $\Lhe(M)$ functions behaving similarly to the spline
curve shown in \reffig{fig_lum_dist}. This behaviour is also the
expected one since stars in the transition region, with mass close to
\Mhef, ignite helium at different degrees of electron degeneracy.
These results, and the simple exercise presented in
\reffig{fig_lum_dist}, leads us to conclude that {\em the secondary
clump shall be even more pronounced than found in models with
0.1~\Msun\ resolution}.

Apart from the mass resolution, another point of concern is the
internal consistency of the stellar tracks computed. In the present
models, the core mass at the ZAHB phase is assumed equal to that at
the onset of the He-flash on the RGB, as well as the envelope chemical
composition. The only effect of the He-flash is assumed to be the
conversion of 5 percent of the helium core mass into carbon. Instead,
tracks computed under the assumption of a constant core mass for all
ZAHB models of low-mass stars would fail to produce the decrease in
the clump luminosity as we approach \Mhef\ (see \reffig{fig_mcore}),
and hence fail to describe the details of the secondary clumps. On the
other hand, models which follow the complete stellar evolution during
the He-flash in a self-consistent way, would probably give an even
better description of the \Mhef\ transition region than the present
ones.

\subsection{Observations of the secondary clump}
\label{sec_observ}

Is the secondary red clump observed in real stellar populations?
Paper~I presents a discussion of the local clump defined by {\em
Hipparcos} data. In that work, the great similarity between the
predicted and observed CMDs immediately suggested the presence of the
secondary clump in the data (see figures~2, 3 and 4 in Paper~I). The
two only visual binaries located in the `secondary clump region' of
the {\em Hipparcos} CMD and with reliable orbital parameters, resulted
to have primary stars with masses consistent with the prediction that
they should be close to $\Mhef\simeq2$~\Msun\ (see also
\refsec{sec_mhef} below).

Bica et al.\ (1998) recently presented CMDs in the Washington system
for several clusters and surrounding fields in the LMC. In two of the
fields in the northest part of this galaxy, close to the clusters
SL~388 and SL~509, a striking secondary clump was observed {\em about
0.45~mag below the dominant intermediate-age clump, and at its blue
side}.  The same feature was present in a third field, SL~769, located
closer to the LMC bar and many degrees away from the above-mentioned
ones.

The authors tentatively suggested that the second clump could be the
signature of a tidal arm or dwarf galaxy located about 10~kpc beyond
the LMC, at a distance similar to the SMC.  Of course, the present
models raise the possibility that the secondary, fainter clumps are
simply caused by a younger stellar population inside the LMC. The
reader is invited to compare the simulations of the LMC clump
presented in the figure~9 of Paper~I (and also in Girardi 1998), with
the observed CMDs shown in figure~4 of Bica et al.\ (1998). These
diagrams are not completely equivalent: the Paper~I models are
presented in the \mi\ vs.\ \vi\ plane, while the latter observations
are presented in the $T_1$ vs.\ $C-T_1$ plane of Washington
photometry. Appart from the use of these different systems, the
modeled clumps are remarkably similar to those observed in the SL~388
and SL~509 fields.

But how could this secondary clump be observed only in some of the LMC
fields? To answer this question, let us recall the conditions for
its appearance in a CMD:
	\benu
	\item
The observed field should contain a significant number of stars about
1~Gyr old, and hence with $\Mhe\sim2$~\Msun, mixed to a population of
older stars.
	\label{item_number}
	\item
The 1~Gyr old population should not be too dispersed in colour due to
its intrinsic metallicity dispersion. The bulk of clump stars should
be more metal-rich than about $Z=0.004$.
	\label{item_metallicity}
	\item
Both main and secondary clumps should not be mixed together due to
effects such as differential reddening, distance dispersions, and
photometric errors.  The combination of these effects should be
producing r.m.s.\ dispersions lower than about 0.2~mag in the apparent
magnitudes of clump stars.
	\label{item_error}
	\eenu

Hubble Space Telescope (HST) observations generally include too few
clump stars to comply with item~\ref{item_number}. On the contrary,
ground-based observations can easily sample large numbers of clump
stars, but in general present the problems referred to in
item~\ref{item_error}. Moreover, many galaxy fields are expected not
to comply with item~\ref{item_metallicity}. Therefore, it is no
surprise that the secondary clumps have not been noticed for so long.
The bimodal clumps observed by Bica et al.\ (1998) may be the
signature of regions with an unusually large population of 1~Gyr old
stars, and/or with unusually small dispersion of reddening. In fact,
the SL~388 and SL~509 fields are located in the outer parts of the LMC
disk, about $5^{\rm o}$ away from the bar. In these regions, the
reddening internal to the LMC should be very small, as well as its
dispersion.

Moreover, the dual clump observed in the field of SL~509 is found to
be present in the cluster itself, with the fainter clump this time
being more pronounced than the bright one (Bica et al.\ 1998). It is
then interesting to consider the age of this cluster.  Its integrated
$UBV$ colours ($\bv=0.73$, $\ub=0.19$; Bica et al.\ 1996) indicate an
$S$-parameter of $41$ (see Elson \& Fall 1985; Girardi et al.\ 1995),
which is equivalent to an age of $\logt=9.22\pm0.23$
($1.66^{+1.15}_{-0.81}$~Gyr), according to the $S$-age calibration of
Girardi et al.\ (1995; eq.\ 5 therein).  On the other hand, Bica et
al.\ (1998) derive an age of 1.2~Gyr for SL~509, based on a
calibration between the magnitude difference from the turn-off to the
red clump, $\delta T_1$, and the age (Geisler et al.\ 1998). In this
age range, however, $\delta T_1$ is expected to become less sensitive
to age due to the leafting of core degeneracy, which causes the clump
to have magnitudes similar to the turn-off. Nonetheless, $\delta T_1$
provides a firm upper limit to the age of the cluster, which we
estimate as being of 1.5~Gyr.  Thus, the age of this cluster is
probably comprised between 0.85 and 1.5~Gyr, which is consistent with
the idea that it contains evolved stars just massive enough for having
ignited helium in non-degenerate conditions. SL~509 is then probably
similar to the cases NGC~2209 and NGC~2190 indicated in
\reffig{fig_corsi}.

We conclude that {the \em Bica et al.\ (1998) data provides consistent
evidence for the existence of the secondary clump in the LMC field
population.} Of course, the question arises whether the same feature
can be identified in other previously published CMDs as well. Some
CMDs indeed present clumps with faint extensions which are compatible
with the characteristics of the secondary clump predicted by theory:
e.g.\ the HST data of Holtzmann et al.\ (1997, their figure~2),
Sarajedini (1998, their field WF2 in figure~4) and Smecker-Hane et
al.\ (1998), the $VR$ photometry of Beaulieu \& Sackett (1998, their
least crowded field F2), and the $VI$ of Stappers et al.\ (1997, their
figure~1). In none of these cases however the secondary clump is as
striking to the eye as in Bica et al.\ (1998) data. A particularly
unfortunate circunstance is that most authors (including those listed
above) present CMDs in which the $V$ or F555W magnitudes are in the
abcissa, whereas the secondary clump would be better separated from
the main one if redder pass-bands were used instead.

	\begin{figure}
	\psfig{file=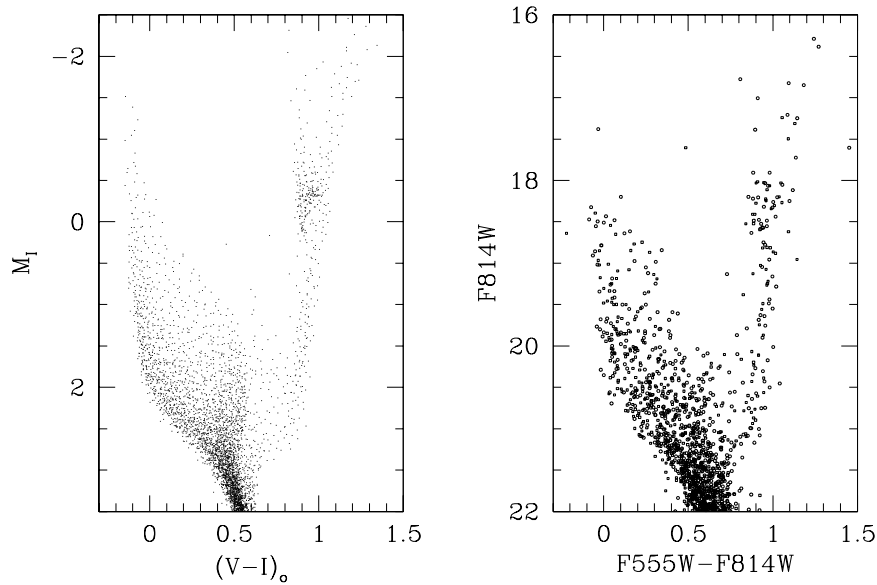,width=12cm}
        \caption{Comparison between a synthetic CMD (left panel) and
the HST data for the LMC field (right panel).  The simulation contains
5000 stars, distributed according to the model described in the text.
The data corresponds to the fields F1 (squares) and F2 (circles)
studied by Holtzmann et al.\ (1997) and Geha et al.\ (1998). }
	\label{fig_mapa_lmc}
	\end{figure} 

In \reffig{fig_mapa_lmc} we present a comparison between our models
and the Geha et al.\ (1998) HST WFPC2 data for two particular LMC
fields (see also Holtzmann et al.\ 1997).  The right panel presents
the F814W vs.\ F555W$-$F814W CMD from the combined LMC fields referred
to as F1 and F2 in Geha et al.\ (1997, see their table~1). We recall
that the HST filters F555W and F814W are roughly equivalent to the $V$
and $I$ ones.

The left panel presents the synthetic \mi\ vs.\ \vi\ CMD for a model
produced by the sum of two different components: (i) the first one has
a constant SFR over the age interval from 0.1 to 3~Gyr ago, with a
gaussian distribution of metallicities of mean $Z=0.008$ and
dispersion $\sigma_Z=0.002$; whereas (ii) the second one presents a 5
times lower SFR from 3 to 10~Gyr ago, with a mean $Z=0.004$ and
dispersion $\sigma_Z=0.002$.  This particular model is by no means a
careful attempt in order to describe the data; it is just meant to be
a synthetic CMD which reasonably resembles the observed one,
especially in the region of the subgiants and red giants.

What is most remarkable in this figure is the similarity between the
clumps in both panels, which clearly span a range in magnitude of
about 0.6~mag. In the simulation, this width is caused mainly by the
presence of the secondary clump. In the data, the photometric errors
at the clump level are simply neglegible (of $\sim0.03$~mag, see
figure~3 in Holtzmann et al.\ 1997), which is probably also the case
for the dispersion in the internal reddening. Therefore, the observed
clump width cannot be accounted by these factors. Instead, we regard
{\em the intrinsic clump structure, as suggested by the models, as the
simplest explanation for the large clump width observed in these LMC
fields}.

Mermilliod et al.\ (1998) recently reported another very interesting
observation: a dual clump is present among the members of the open
cluster NGC~752. The brighter clump is composed of 8 stars, whereas
the fainter and bluer one contains 3 or 4 stars. Moreover, this
cluster has an age close to the limit for having non-degenerate helium
ignition (see e.g.\ Daniel et al.\ 1994; Carraro \& Chiosi
1994). Therefore, we face the interesting possibility of observing a
cluster at the right age to contain both the clump stars which went,
and those which did not went, through the helium flash.  We remark
that the probability of finding such a cluster is somewhat low,
because of the narrow range of initial masses typically covered by the
clump stars in individual clusters. On the contrary, dual clumps
should be frequent in galaxy fields.

\subsection{The secondary clump as a tracer of the star formation
history}
\label{sec_sfr}

The secondary red clump is a CMD feature sensitive to the history of
star formation in the parent galaxy field. A good example of this may
be found in the LMC. Indeed, most studies of the field population
indicate that the LMC disc is relatively young on average, having
formed a large fraction of its stars (if not most) in the last 4~Gyr
or so (see e.g.\ Bertelli et al.\ 1992; Vallenari et al.\ 1996;
Holtzmann et al.\ 1997; Geha et al.\ 1998). This circunstance may
contribute to make the secondary clump in this galaxy particularly
strong if compared to the older, main clump.

Models like the one presented in \reffig{fig_mapa_lmc} suggest that
the number of stars in the secondary clump feature should be simply
proportional to the SFR at the age of 1~Gyr. In fact, this CMD feature
is expected to be an excellent tracer of the spatial distribution of
star formation in the LMC at $\sim1$~Gyr ago.  Hence, {\em it would be
extremely interesting to check the available photometric data} (e.g.\
in the EROS, MACHO, OGLE, and Magellanic Cloud Photometric Survey
databases) {\em in order to identify the secondary clump in different
fields of the LMC}, and hence possibly to map its SFR at 1~Gyr ago.
Of course, the same comments apply to all Local Group galaxies for
which high-quality and uncrowded stellar photometry is feasible.

The number ratio between stars belonging to the main and secondary
clumps, is affected by both the IMF and the particular details of the
SFR history over a much longer age interval, from 1~Gyr to
$12-15$~Gyr. Thus, this number ratio does not provide an unique
interpretation in terms of the SFR. The study of the complete CMD
would be necessary to this aim. We remark, however, that {\em the
observed clump structure can provide additional constraints (which
have so far been neglected) for the derivation of the SFR history in
galaxy fields}.

\subsection{Measuring \Mhef\ with stars in the secondary clump}
\label{sec_mhef}

Figure \ref{fig_lum_dist} shows that stars in the secondary clump
should have masses close to \Mhef. More specifically, for a single
value of metallicity, stars more than 0.2~mag below the main red clump
have masses contained within the relatively narrow interval of
$[\Mhef+0.2\Msun, \Mhef-0.1\Msun]$.

The limiting mass \Mhef\ (\refsec{sec_theory}) is sensitive to the
adopted chemical composition, and to the treatment of the convective
boundaries of the core during the main sequence phase. The upper panel
of \reffig{fig_mcore} and \reftab{tab_mhef} illustrate how its value
gets lower for lower values of metal and helium content. The values we
get (e.g.\ $\Mhef=2.0$~\Msun\ for $Z=0.019$) are typical of models
computed with a moderate amount of convective overshooting. We
computed an additional set of $Z=0.019$ tracks adopting the classical
Schwarzschild criterion for the boundary of convective regions,
obtaining $\Mhef=2.4$~\Msun. Our values agree, to within 0.1~\Msun, to
those found by different authors in models of similar characteristics
(Sweigart et al.\ 1990; Maeder \& Meynet 1989; Bertelli et al.\ 1994,
and references therein).

These considerations led us to suggest, in Paper~I, {\em the use of
secondary clump stars located in binary systems in order to
observationally constrain \Mhef}. The mass information from the binary
orbit, together with some measure of the stellar metallicity, could
allow to get a direct measure of \Mhef, and hence constraints on the
efficiency of convective overshooting in stellar cores.

	\begin{figure}
	\psfig{file=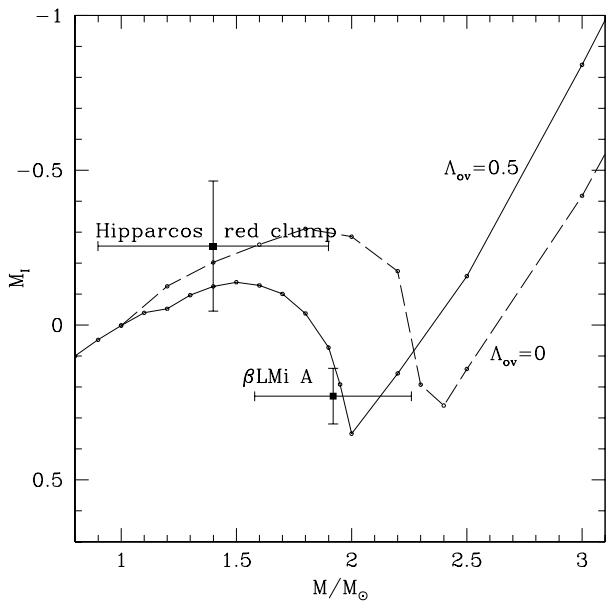,width=8.3cm}
        \caption{The $I$-band absolute magnitude of clump stars as a
function of their mass, for solar metallicity. The lines delimit the
lower boundary of the clump as predicted by models with and without
overshooting (solid and dashed lines, respectively).  The mean
position of the Hipparcos clump, and of the $\beta\,$LMi primary star,
are also presented. The position of $\beta\,$LMi in this diagram
indicates a star just massive enough for having ignited He in
non-degenerate conditions, and would favour the case of moderate
convective overshooting.}
	\label{fig_binary}
	\end{figure} 

Figure~\ref{fig_binary} illustrates the most favorable case we find in
the {\em Hipparcos} catalog of binary stars (ESA 1997). It regards the
visual binary $\beta$~LMi (HD~90537), which has well-measured orbital
parameters (Heintz 1982) and an uncertainty in {\em Hipparcos}
parallax of 4 percent. The primary star, with a derived mass of
$1.92\pm0.34$~\Msun, probably belongs to the secondary clump defined
in the CMD from {\em Hipparcos} (see Paper~I). Its spectrum indicates
$\feh=0.0$ (McWilliam 1990), i.e.\ solar metallicity. In the figure,
we show the \mi\ absolute magnitude as a function of mass for the
lowest-luminosity CHeB models of $Z=0.019$ (cf.\ \reffig{fig_mcore}),
for both the classical ($\Lambda_{\rm c}=0$) and moderate overshooting
($\Lambda_{\rm c}=0.5$; see Alongi et al.\ 1993; Chiosi et al.\ 1992,
and references therein) cases. The mean position and dispersion of the
{\em Hipparcos} red clump is indicated by error bars. Similarly, we
plot the probable position of the primary star $\beta$~LMi~A with the
corresponding standard errors. Assuming that this star is in the stage
of CHeB (as its position in the {\em Hipparcos} CMD indicates), its
measured mass would be consistent with a value of
$\Mhef\simeq1.9$~\Msun, and would favour the case of moderate
convective overshooting as indicated by the $\Lambda_{\rm c}=0.5$
line.

Of course, no firm conclusions about the efficiency of convective
overshooting can be taken from a single star, for which the parameters
are not known with the desired accuracy. We have further checked the
data for well-studied visual binaries in the solar vicinity, finding
no better candidate to belong to the secondary clump. Future or
alternative data may improve upon this result.

\subsection{The red clump as a distance indicator}
\label{sec_distance}

There are two ways in which the red clump can be used as a distance
indicator. The first one has been commented on in \refsec{sec_intro},
and regards the {\em absolute distance determinations} of Local Group
galaxies. It relies on the clump absolute magnitude \micl\ as measured
from local stars in the {\em Hipparcos} database, and on the
assumption that it is constant for different galaxies. The result on a
very short distance to the Magellanic Clouds (Udalski et al.\ 1998;
Stanek et al.\ 1998) caused the suspicion that systematic effects
could be present in the method. Cole (1998) and Paper~I identified the
dependence of the clump absolute magnitude with age and metallicity,
suggested by the models, as the main responsible for the discrepant
results for the Clouds. Additionally, Gratton, Carretta \& Clementini
(1998) call attention to the large values of the absorption $A_I$ to
the Clouds used by Udalski et al.\ (1998) and Stanek et al.\
(1998). If more conservative values of $A_I$ and the metallicity-age
corrections to \micl\ are adopted, the LMC distance as derived from
the red clump can yet be reconciled with the more traditional values
of $\dmo=18.5\pm0.1$~mag. Clearly, the red clump stars alone do not
provide a definitive value to the LMC distance modulus.

Comparing the clump and RR Lyrae data for fields in the LMC, SMC, and
Carina dwarf galaxy, Udalski (1998a) try to access the dependence of
\micl\ on metallicity, concluding that it varies very little, i.e.\
$\micl=(0.09\pm0.03)\, \feh + {\rm constant}$. However, this
conclusion largely holds on the assumption that the RR Lyrae in the
observed fields provide accurate standard candles, and that the RR
Lyrae absolute magnitudes vary with metallicity as $M_V =
(0.18\pm0.03)\,\feh + {\rm constant}$. At least the second point may
be considered as controversial (see eg.\ Gratton et al.\ 1998;
Popowski \& Gould 1998).

Udalski (1998b) also attempt to derive the dependence of \micl\ on the
age of Magellanic Cloud clusters, limiting the analysis to the age
interval from 2 to 15~Gyr. His data has been discussed in
\refsec{sec_clump_brightness}. Considering the undertainties on the
reddening and distances for individual star clusters, Udalski's
(1998b) data is consistent with the behaviour of clump magnitudes
predicted by models (\reffig{fig_udalski}). Udalski (1998ab) instead
refers to the models as `involving many difficult to verify and often
controversial assumptions like mass-loss, helium content, star
formation rate, etc'. We would like to stress that uncertainties in
the mass-loss rates play a negligible role in determining the clump
structure in the local {\em Hipparcos} sample and in the LMC. This is
so because all stars with $M>1.5$~\Msun\ suffer from negligible
mass-loss along the RGB. Only the reddest tail of these clumps may be
slightly affected by mass-loss, which would however not change
appreciably the derived clump mean magnitudes.  On the other hand, the
helium-to-metal content relation adopted in the present models,
$Y(Z)\simeq0.23+2.25\,Z$ (see \reftab{tab_mhef} for the precise $Y$
values), represents a fairly conservative one. Independent of these
assumptions, the models are quite clear in indicating a dependence of
the clump $I$-band absolute magnitude on metallicity, a point which is
crucial to the method of distance determinations based on clump stars.

One of the main uncertainties in the models, for instance, may be
identified in the adopted transformations from the theoretical to
observational planes (bolometric corrections and \Teff--colour
relations, e.g.\ Kurucz 1992).  Inadequacies in these tranformations
are able to systematically shift models in their colours and absolute
magnitudes. But since these are probably {\em mostly systematic}
errors, they cannot significantly affect the relative values between
these quantities as derived from models. The metallicity dependence of
\micl, for instance, largely reflects the dependence of \logL\ on this
parameter (see \reffig{fig_mcore}), and not the bolometric
corrections.

The magnitude distribution of the red clump has also been used as a
{\em relative distance indicator}.  Hatzidimitriou \& Hawkins (1989),
Gardiner \& Hawkins (1991) and Gardiner \& Hatzidimitriou (1992) have
used the width of the luminosity function of red clump stars in
different regions of the SMC, in order to map its depth along the
line-of-sight.  For some regions, the clumps were observed to be very
narrow in magnitude, with standard widths of about
$\sigma=0.1$~mag. On the other hand, other regions presented clump
widths of up to $\sigma=0.5$~mag. These broad clumps were interpreted
as being the signature of different stellar populations located at
different distances. This assumption lead to the derivation of depths
of the SMC populations along the line-of-sight, which range from $4-6$
to $12-16$~Kpc in different regions of the SMC (Gardiner \& Hawkins
1991).

The present models instead indicate that the red clump is not
intrinsically narrow in all situations. In the particular cases of
metal-poor galaxy fields with a constant SFR, for instance, the red
clump is expected to have an intrinsic width of at least
$\sigma=0.2$~mag (see e.g.\ the left panels of \reffig{fig_secclump}).
Slightly larger intrinsic widths may be present if the star formation
has been particularly intense at 1~Gyr ago (see e.g.\
\reffig{fig_mapa_lmc}), due to the appearance of the secondary
clump. In this case, the increased clump width is expected to
correlate with the presence of a larger population of main sequence
stars nearly as bright as (or even brighter than) the clump level in
the CMD. Moreover, additional broadening of the clump is expected when
a field presents a significant metallicity dispersion.

Therefore, part of the observed clump width in the SMC may possibly be
interpreted as intrinsic variations of the clump absolute magnitude,
rather than the inferred depth structures. The same comment apply as
well to the LMC population. A re-evaluation of the Magellanic Clouds
clump data may be of interest in order to clarify these points.  We
remark, however, that the kinematical data of Hatzidimitriou, Cannon
\& Hawkins (1993) seem to confirm the presence of different structures
along the SMC line-of-sight, at least in the areas of maximum inferred
depth.

Another example of the use of the clump as a relative distance
indicator is the work by Zaritsky \& Lin (1997). Having identified a
brighter extension of the LMC clump (up to $\sim1$~mag wide), they
suggested it should be caused by an intervening stellar population
located between us and that galaxy.  This would have important
implications to the interpretation of the observed rate of
micro-lensing events towards the LMC. However, Beaulieu \& Sackett
(1998) and Ibata, Lewis \& Beaulieu (1998) convincingly argued that
this `vertical red clump' should rather be caused by the youngest red
clump stars (see also Sects.\ \ref{sec_subclump} and \ref{sec_observ}
above). This further illustrates that the red clump should be used
with care in distance determinations, and that synthetic CMDs may be
of invaluable help in the analysis of similar data. Moreover, it
evidences that intermediate-mass stars cannot be neglected in the
modelling of the red clump.

\section{Concluding remarks}
\label{sec_conclu}

The main results of the present work may be summarized as follows:
	\benu
	\item
Present data about the clump magnitude in Magellanic Cloud clusters
are consistent with the behaviour predicted by stellar evolutionary
models.  Particularly interesting is the presence of a minimum in the
luminosity of clump stars for an age of about $\sim1$~Gyr, clearly
evidenced by the Corsi et al.\ (1994) data, which corresponds to the
maximum age for stars which ignite helium under non-degenerate
condictions. The presence of this luminosity minimum is therefore a
solid theoretical prediction.  This minimum occurs for stars with mass
$\Mhef=2.0$~\Msun\ if the metallicity is about solar and moderate
convective overshooting is assumed. In models without convective
overshooting, it would rather correspond to ages of about 0.5~Gyr, and
stellar masses of about 2.4~\Msun.
	\item
The same stellar models are used to generate synthetic CMDs for
different galaxy fields. It turns out that these models predict the
presence of a secondary red clump in the CMD, which is located about
0.4~mag below the main clump, and slightly to its blue. This feature
is expected to be present in fields which formed stars $\sim1$~Gyr
ago, and in which the typical metallicities are higher than about
$Z=0.004$.
	\item
In order to describe well this feature in synthetic CMDs, stellar
evolutionary tracks should be computed with a fine mass resolution in
the complete relevant mass interval, i.e.\ from the lowest possible
mass of clump stars, up to about $\Mhef+0.5$~\Msun. In the vicinity of
\Mhef, a mass resolution of at least $\Delta M=0.1$~\Msun\ is
required. Moreover, this work evidences that the mass interval of
$\Mhe\ga\Mhef$ may represent a significant fraction of the clump
stars. Therefore, it cannot be neglected in detailed studies of the
red clump, especially in regard to galaxies that have kept forming
stars in the last few gigayears of their evolution.
	\item
Secondary clumps consistent with the model predictions are present in
{\em Hipparcos} CMD for nearby stars, and in some LMC fields observed
by Bica et al.\ (1998). Similar features are present, although not too
clear, in LMC data presented by several other authors. Secondary
clumps may be missing or hidden in the CMDs of other galaxy fields due
either to the poor number statistics, or to a large dispersion in
reddening and metallicity, or to photometric errors, or to the
absence/scarcity of 1~Gyr old populations. Data for fields in the
Magellanic Clouds and other Local Group galaxies should be carefully
re-investigated in the light of these results. In fact, this fine
structure of the red clump may provide important constraints to the
star formation history of nearby galaxies, as well as to the
distribution of their stars along the line-of-sight.
	\eenu

\section*{Acknowledgments}
Thanks are due to E.\ Bica, A.\ Bressan, C.\ Chiosi and A.\ Weiss for
their useful comments and suggestions. A.A.\ Cole and L.\ Gardiner are
acknowledged for calling our attention to some important observational
work here mentioned (regarding NGC~752 and the SMC,
respectively). A.A.\ Cole, J.S.\ Gallagher and J.A.\ Holtzman kindly
provided the data for \reffig{fig_mapa_lmc}. We made use of the Simbad
database, mantained by the CDS, Strasbourg. This work is funded by the
Alexander von Humboldt-Stiftung.

\label{lastpage}


\begin{thebibliography}{99}
\bibitem{} Alongi M., Bertelli G., Bressan A., Chiosi C., Fagotto F.,
	Greggio L., Nasi E., 1993, A\&AS 97, 851
\bibitem{} Alves D.R., Basu A., Cook K.H., et al., 1998, in New Views
	of the Magellanic Clouds, IAU Symp.\ 190, eds.\ Chu Y.-H.,
	Suntzeff N., Hesser J., Bohlender D., in press
	(astro-ph/9810221).
\bibitem{} Beaulieu J.-P., Sackett P.D., 1998, AJ 116, 209
\bibitem{} Becker S.A., Iben I.\ Jr., 1980, ApJ 237, 111
\bibitem{} Bertelli G., Mateo M., Chiosi C., Bressan A., 1992, ApJ
	388, 400
\bibitem{} Bertelli G., Bressan A., Chiosi C., Fagotto F, Nasi E., 
      1994, A\&AS, 106, 275
\bibitem{} Bica E., Clari\'a J.J., Dottori H., Santos Jr.\ J.F.C., Piatti
	A.E., 1996, ApJS 102, 57
\bibitem{} Bica E., Geisler D., Dottori H., Clari\'a J.J., Piatti
	A.E., Santos Jr.\ J.F.C., 1998, AJ 116, 723
\bibitem{} Cannon R.D., 1970, MNRAS 150, 111
\bibitem{} Carraro G., Chiosi C., 1994, A\&A 287, 761
\bibitem{} Castellani V., Chieffi A., Straniero O., 1992, ApJS 78, 517
\bibitem{} Chiosi C., Bertelli G., Bressan A., 1992, ARA\&A 30, 235
\bibitem{} Cole A.A., 1998, ApJ 500, L137
\bibitem{} Corsi C.E., Buonanno R., Fusi Pecci F., Ferraro F.R., Testa
	V., Greggio L., 1994, MNRAS 271, 385
\bibitem{} Da Costa G., 1991, in The Magellanic Clouds, IAU Symp.\
	148, eds.\ Haynes R., Milne D., Dordrecht: Kluwer, p.\ 183
\bibitem{} Da Costa G., Hatzidimitriou D., 1998, AJ 115, 1934
\bibitem{} Daniel S.A., Latham D.W., Mathieu R.D., Twarog B.A., 1994,
	PASP 106, 281
\bibitem{} Elson R.A.W., Fall S.M., 1985, ApJ 299, 211
\bibitem{} ESA, 1997, The Hipparcos and Tycho Catalogues, ESA SP-1200
\bibitem{} Faulkner D.J., Cannon R.D., 1973, ApJ 180, 435
\bibitem{} Gardiner L.T., Hawkins M.R.S., 1991, MNRAS 251, 174
\bibitem{} Gardiner L.T., Hatzidimitriou D., 1992, MNRAS 257, 195
\bibitem{} Geisler D., Bica E., Dottori H., Clari\'a J.J., Piatti
	A.E., Santos Jr.\ J.F.C., 1997, AJ 114, 1920
\bibitem{} Geha M.C., Holtzman J.A., Mould J.R., et al., 1998,
	 115, 1045
\bibitem{} Girardi L., Chiosi C., Bertelli G., Bressan A., 1995, 
	A\&A 298, 87
\bibitem{} Girardi L., 1998, in New Views of the Magellanic Clouds,
	IAU Symp.\ 190, eds.\ Chu Y.-H., Suntzeff N., Hesser J.,
	Bohlender D., in press.
\bibitem{} Girardi L., Bertelli G., 1998, MNRAS 300, 533
\bibitem{} Girardi L., Groenewegen M.A.T., Weiss A., Salaris M., 1998,
	 MNRAS 301, 149
\bibitem{} Girardi L., Bressan A., Bertelli G., Chiosi C., 1999, 
	in preparation
\bibitem{} Gratton R., Fusi Pecci F., Carretta E., Clementini G.,
	Corsi C.E., Lattanzi M., 1997, ApJ 491, 749
\bibitem{} Gratton R., Carretta E., Clementini G., 1998, in
	Post-Hipparcos Standard Candles, eds.\ A.\ Heck \& F.\ Caputo, 
	Kluwer: Dordrecht, in press.
\bibitem{} Hatzidimitriou D., 1991, MNRAS 251, 545
\bibitem{} Hatzidimitriou D., Hawkins M.R.S., 1989, MNRAS 241, 667
\bibitem{} Hatzidimitriou D., Cannon R.D., Hawkins M.R.S., 1993, MNRAS
	261, 873
\bibitem{} Heintz W.D., 1982, PASP 94, 705
\bibitem{} Holtzman J.A., Mould J.R., Gallagher J.S., et al., 1997, AJ
	113, 656
\bibitem{} Ibata R.A., Lewis G.F., Beaulieu J.-P., 1998, ApJ 509, L29
\bibitem{} Kurucz R.L., 1992, in The Stellar Populations of Galaxies, 
        eds.\ B.\ Barbuy and A.\ Renzini, Dordrecht: Kluwer, p.\ 225
\bibitem{} Maeder A., Meynet G., 1989, A\&A 1989 210, 155
\bibitem{} Mateo M., Hodge P., Schommer R.A., 1986, ApJ 311, 113
\bibitem{} McWilliam A., 1990, ApJS 74, 1075
\bibitem{} Mermilliod J.-C., Mathieu R.D., Latham D.W., Mayor M., 1998,
	A\&A 339, 423
\bibitem{} Olszewski E.W, Schommer R.A., Suntzeff N.B., Harris H.C.,
	1991, AJ 101, 515
\bibitem{} Olszewski E.W, Suntzeff N.B., Mateo M., 1996, ARA\&A 34, 511 
\bibitem{} Paczy\'nski B., Stanek K.Z., 1998, ApJ 494, L219 
\bibitem{} Perryman M.A.C., et al., 1997, A\&A 323, L49
\bibitem{} Popowski P., Gould A., 1998, in Post-Hipparcos Standard
	Candles, eds.\ A.\ Heck \& F.\ Caputo, Kluwer: Dordrecht, in
	press.
\bibitem{} Reimers D., 1975, Mem.\ Soc.\ R.\ Sci.\ Li\`ege, 
        ser.\ 6, vol.\ 8, p.\ 369 
\bibitem{} Renzini A., Fusi Pecci F., 1988, ARA\&A 26, 199
\bibitem{} Salaris M., Weiss A., 1997, A\&A 327, 107
\bibitem{} Salaris M., Weiss A., 1998, A\&A 335, 943
\bibitem{} Sarajedini A., 1998, AJ 116, 738
\bibitem{} Searle L., Sargent W.L.W., 1972, ApJ 173, 25
\bibitem{} Seidel E., Demarque P., Weinberg D., 1987, ApJS 63, 917
\bibitem{} Seidel E., Da Costa G.S., Demarque P., 1987, ApJ 313, 192
\bibitem{} Smecker-Hane T., Gallagher J.S., Cole A.A., Tolstoy E.,
	Stetson P.B., 1998, in New Views of the Magellanic Clouds, IAU
	Symp.\ 190, eds.\ Chu Y.-H., Suntzeff N., Hesser J., Bohlender
	D., in press.
\bibitem{} Stanek K.Z., Garnavich P.M., 1998, ApJ 503, L131
\bibitem{} Stanek K.Z., Zaritsky D., Harris J., 1998, ApJ 500, L141
\bibitem{} Stappers B.W., Mould J.R., Sebo K.M., et al., 1997, PASP
	109, 292 
\bibitem{} Sweigart A.V., Greggio L., Renzini A., 1989, ApJS 69, 911
\bibitem{} Sweigart A.V., Greggio L., Renzini A., 1990, ApJ 364, 527
\bibitem{} Tolstoy E., Gallagher J.S., Cole A.A., Hoessel J.G., Saha
	A., Dohm-Palmer R., Skillman E., Mateo M., Hurley-Keller D.,
	1998, AJ 116, 1244
\bibitem{} Udalski A., Szyma\'nski M., Kubiak M., Pietrzy\'nski G.,
	Wo\'zniak P., $\dot{\rm Z}$ebru\'n K, 1998, Acta Astr.\ 48, 1
\bibitem{} Udalski A., 1998a, Acta Astr.\ 48, 113
\bibitem{} Udalski A., 1998b, Acta Astr.\ 48, 383
\bibitem{} Vallenari A., Chiosi C., Bertelli G., Aparicio A., Ortolani
	S., 1996 A\&A 309, 367
\bibitem{} Zaritsky D., Lin D.N.C, 1997, AJ 114, 2545
\bibitem{} Zinn R., West M.J., 1984, ApJS 55, 45

\end{thebibliography}
\end{document}